# Magnetic field driven quantum criticality in antiferromagnetic $CePtIn_4$

Debarchan Das, Daniel Gnida, Piotr Wiśniewski, and Dariusz Kaczorowski*
*Institute of Low Temperature and Structure Research, Polish Academy of Sciences, Wrocław 50-950, Poland*

* Corresponding author electronic address: d.kaczorowski@intibs.pl

Physics of quantum critical point is one of the most perplexing topics in current condensed-matter physics. Its conclusive understanding is forestalled by the scarcity of experimental systems displaying novel aspects of quantum criticality. We present a comprehensive experimental evidence of a magnetic field tuned tricritical point separating paramagnetic, antiferromagnetic and metamagnetic phases in novel $CePtIn_4$. Analyzing field variations of its magnetic susceptibility, magnetoresistance and specific heat at very low temperatures we trace modifications of antiferromagnetic structure of the compound. Upon applying magnetic field of increasing strength, the system undergoes metamagnetic transitions which persist down to the lowest temperature investigated, exhibiting first-order-like boundaries separating magnetic phases. This yields a unique phase diagram where the second-order phase transition line terminates at a tricritical point followed by two first-order lines reaching quantum critical end points as $T \rightarrow 0$. Our findings demonstrate that $CePtIn_4$ provides innovative perspective for studies of quantum criticality.

## Introduction

For past two decades understanding quantum critical point (QCP) associated with a quantum phase transition (QPT) in strongly correlated electron systems has become one of the outstanding challenges in modern physics due to its affluent and fairly enigmatic physics content. QPT leads the system through a precarious point of instability (second-order) at $T = 0$ separating two stable phases (e.g. magnetic and non-magnetic), known as QCP [1, 2]. Interestingly, QCP can also be achieved when the critical end-point terminating a line of first-order transitions (without any broken symmetry) is suppressed to $T = 0$ [3, 4]. In contrast to a conventional classical phase transition which is driven by temperature, QPT and thus QCP is tuned by non-thermal parameters such as external pressure, chemical pressure (doping) or magnetic field [1, 5, 6]. However, there are some difficulties while employing pressure and chemical doping as tuning parameters: it is difficult to carry out various thermodynamic measurements a very low temperatures and under high pressure, whereas alloying tends to introduce disorder-driven effects which can not be separated from the quantum criticality of the translational invariant system. In this regards, magnetic field appears to be of particular usefulness as it can reversibly and continuously tune the system towards QCP without encountering aforementioned complexities [7]. Even though there are quite a few systems described in the literature manifesting magnetic field tuned QCP such as $YbRh_2Si_2$ [8-9], $Sr_3Ru_2O_7$ [10-12] etc. [15-18], yet there is no universality in their behavior. This foregrounds the necessity of novel systems to fully understand magnetic field tuned QCP. Thus, the search for novel systems manifesting QCP has recently become an exigent goal. In this endeavor, Ce based compounds turn out to be quite promising due to unstable nature of 4*f* orbitals which allows relatively easy tuning of magnetic properties leading to QCP. Recent reports on close proximity of superconductivity and antiferromagnetism (AFM) in heavy-fermion systems having chemical formula $Ce_nTm_mIn_{3n+2m}$ (where Tm stands for a d-electron transition metal) [19-24], have ignited a new thrill amongst the condensed matter research community. A distinguished member of this family is $CeCoIn_5$ (with n=1 and m= 1) which is a heavy-fermion superconductor (HFSC) with $T_c$ =2.3 K [19]. $CeCoIn_5$ gained profound interest due to associated quantum criticality [15, 25, 26] and formation of Q-phase [27]. $Ce_2PdIn_8$ (n=2 and m= 1) is another example of HFSC having $T_c$ = 0.7 K which manifests a non-Fermi liquid type normal state behavior dominated by AFM spin fluctuations with possible Fulde-Ferrell-Larkin-Ovchinnikov character [17, 28, 29]. In this context it is noteworthy to mention $Ce_3TmIn_{11}$ (n=3, m=1 and Tm=Pt, Pd) in particular, as it turns out to be quite a unique example where superconductivity coexists with AFM ordered state with a pressure induced quantum criticality [23, 24]. Moreover, a field-induced phase diagram of this system is very complex [30, 31]. These unusual observations pave the way to look for new CeTmIn (Tm= Pt and Pd) based ternary compounds including hypothetical$CePtIn_5$ and $CePdIn_5$.

In our ongoing quest to synthesize $CePtIn_5$, we recently discovered a novel phase $CePtIn_4$ grown in a platelet like morphology crystallizing in an orthorhombic structure (Cmcm, Space group No. 63) [32]. Surprisingly, in contrast to needle shape crystals reported by Bławat *et al.* [33], platelet like crystals exhibit slightly elevated $T_N$ = 2.3 K [32] due to the ordering of $Ce^{3+}$ moments. This enhancement in $T_N$ can be due to enlarged unit cell volume. Preliminary results on these novel crystals tempted us to explore the possible existence of QCP in this system by suppressing its AFM order. Here, we report the results of our magnetic, electrical transport and thermodynamic measurements on $CePtIn_4$ accentuating a perplexing magnetic phase diagram which exposes metamagnetic quantum critical end points. This unique and remarkable finding expands the frontier of this new class of QCP generating an opportunity to exploit more experimental and theoretical understanding of associated perplexing physics.

**Results and Discussion**

Figure 1 (a)-(d) illustrates temperature dependence of magnetic susceptibility (*M/H*) with field applied along crystallographic *b*-axis. AFM nature of the ordering is manifested by a prominent sharp peak in *M=H* at Néel temperature (Fig. 1a). Notably, it is quite evident from Fig. 1a, that in addition to the AFM transition at $T_N$ =2.2 K, there is an upturn at ~1 K which appears to be strongly influenced by the applied field. In 1.4 T this upturn becomes distinctly visible and takes a very pronounced shape (Fig. 1b) signaling a phase transition in the system. In this context it must be emphasized that $CePtIn_4$ has only one Ce position in its crystal structure. Thus, this second anomaly must be associated with the same Ce moments which order AFM at the first place. Surprisingly, with further increase in applied field (at 1.7 T, Fig. 1c) we observed single transition. At 5 T, *M/H* (Fig. 1d) reveals a saturation behavior evincing the complete polarization of Ce moments. It is to be noted that zero-field cooled (ZFC) and field cooled (FC) data taken at these fields do not show any characteristic difference originating from domain effect which in turn discards the possibility of ferromagnetic component within this second transition. It can be engendered due to reorientation of magnetic moments forming a novel magnetic order

In Fig. 1(e), we present few isothermal magnetization curves obtained in the AFM ordered state with applied field parallel to the crystallographic *b*-axis (see Fig S1a in Supplementary Material [34] for all isotherms). At very low temperatures, (0.5 K $\leq T \leq$ 0.8) ***M(H)*** curves display two prominent step-like transitions. These distinct features in ***M(H)*** curves unambiguously demonstrate metamagnetic transitions (MMT) in the system. With increasing temperature, these MMT are no longer so prominently visible in ***M(H)*** curve, even though, a clear change in the slope is visible for each isotherm. Furthermore, at sufficiently high fields all magnetization curves tend to saturate at a value ~1.31 $\mu_B$ which is much smaller than that expected for fully ordered magnetic moment of $Ce^{3+}$ ions. This reduced value corresponds very well with a doublet ground state of $^2F_{5}=2$ multiplet split in an orthorhombic crystalline electric field potential. Figure 2(a) depicts the field dependence of differential susceptibility, d*M*/d*H*, for few chosen temperatures (for all temperatures see Fig. S1b in Supplementary Material [34]). Here, the maxima in d*M*/d*H* vs *H* correspond to MMT. The peaks in d*M*/d*H* become sharper and increase in height as the temperature is lowered, suggesting that quantum fluctuations play a key role in governing the breadth of field-driven transition as $T \rightarrow 0$. One crucial observation that demands further attention is the temperature dependence of the critical fields associated with these MMTs. While the position of first MMT (low field transition) is hardly altered by the temperature variation, the high field MMT appears to be very sensitive to the temperature. In Fig. 2(b) we have plotted the maximum value of the differential susceptibility, $dM/dH|_{max}$, as a function of temperature. At elevated temperatures, it tends to diverge with temperature as power law i.e. $dM/dH|_{max} \sim T^{-n}$ (with n = 3.1, see red dashed line). This strongly indicates involvement of long-wavelength quantum critical fluctuations over the *H*-*T* phase diagram in the vicinity of critical field [35]. At lower temperature, $dM/dH|_{max}$ exhibits a saturation behavior which signals a broadening of the AFM transition. Such broadening of transition can be caused either by disorder effect or alternatively by thermal or quantum fluctuations. Given the experimental evidences as presented in the following section, quantum fluctuations are most likely to play a crucial role in governing this behavior.

The temperature dependence of electrical resistivity ρ(*T*) of $CePtIn_4$ measured in the temperature range 0.05-4 K in various fields applied along *b*-axis, is presented in Fig. 3(a). In zero field, below $T_N$ = 2.3 K, ρ(*T*) undergoes a drastic fall as a consequence of reduction of spin disorder scattering due to AFM-type ordering of Ce-moments. In this regard, it is worthwhile to mention that in the earlier report on needle like $CePtIn_4$ single crystals, no such drop in ρ(*T*) was observed [33]. With increasing field, $T_N$ first shifts towards lower temperature, featuring typical characteristic of AFM order. For $\mu_0 H \geq 2$ T the anomaly in ρ(*T*) starts to broaden. Simple analysis of low temperature data (below $T_N$) considering only electron-electron interaction (i.e. $T^2$ dependency) will be misleading in the present case due to significant contribution from scattering of electrons on AFM spin wave excitations. As the actual magnon dispersion relation is not known for this material, we did not perform any analysis of ρ(*T*) data.

To gain further insight into the anomalous behavior seen in ρ(*T*) vs *T* under different applied fields, we measured magnetoresistance (MR = ρ(*H*)/ ρ(*0*) - 1) of $CePtIn_4$ in transverse geometry i.e. H||*b* and electric current flowing within the crystallographic *ac*- plane. Figure 3(b) presents MR measured at various temperatures below $T_N$. It is evident that in the AFM state, MR is positive which is consistent with the AFM ordering in $CePtIn_4$. Quite convincingly, the MR singularities at $H_c$ manifest MMT in $CePtIn_4$. In Fig. 3(c) we presented MR data measured at 0.1 K in field up and down (indicated by arrows in the figure) cycle. A clear hysteresis is visible between increasing and decreasing field sweeps evidencing first order nature of MMT. A rigorous inspection of the MR data as presented in Fig. 3(d) (for such magnified view of all temperature see Fig. S2 in Supplementary Material [34]) reveals an extraordinary feature that well below $T_N$ MR exhibits two successive MMT over a very short range (1.39 T $\leq \mu_0 H \leq$ 1.75 T) of applied field. With increasing temperature two distinct transitions converge to a single one. This behavior can be well visualized in a color contour plot (see Fig. 3e). These findings are in excellent agreement with that witnessed in magnetic isotherms (see Fig. 1e). Another observation drawn from MR data features multiple switching between positive and negative MR values leading to a minimum in the field dependence of MR in paramagnetic state (or field polarized paramagnetic state). This behavior can be reasonably explained considering the interplay between two mechanisms, which contribute to the overall MR with different signs. While, the negative contribution is usually explained by the suppression of spin fluctuations in an external magnetic field larger than critical field [36], the positive contribution can be associated with normal/ordinary MR resulting from the field-induced cyclotron motion of the conduction electrons. Although, one should notice that MR(H) follows a linear dependence without any dendency of a saturation at higher fields, which is quite unusual as regards heavy-fermion AFM systems. It is also worth to emphasize that similar behavior has previously been observed only in a few such systems, including $Ce_3TmIn_{11}$ (Tm= Pt and Pd)[30, 31] and $Ce_2PdGa_{12}$[37], yet its actual origin remains unclear.

In order to elucidate the intrinsic bulk nature of various field induced transitions seen in magnetic, electrical and magnetotransport properties, we measured heat capacity of single crystalline $CePtIn_4$ in different applied fields H||b. Fig. 4(a) represents the temperature variation of the heat capacity over temperature ratio (C/T ) measured for $CePtIn_4$ in magnetic fields up to 2.5 T (for the data up to 7 T see Fig. S3 in Supplementary Material [34]). The bulk AFM ordering is confirmed by the observed $\lambda$ type anomaly in *C/T* in zero applied field. At the lowest temperature investigated *C/T* reaches a significantly large value, signaling the presence of strong electron correlations. In the paramagnetic (PM) state *C/T* follows a -ln*T* dependence for all applied fields characteristic of critical fluctuations in magnetic non-Fermi-liquid (NFL) systems. It is important to note that at zero applied field, apart from the sharp anomaly at $T_N$, there exists another broad anomaly at ~ 1 K (for better visualization see Ref. [32] and Fig. S 4 in Supplementary Material [34]) which is not influenced by the applied field up to 1.3 T. Notably, at 1.4 T, this anomaly becomes considerably sharp and we observe two distinct transitions. This is in accordance with the observations inferred from magnetic and electrical resistivity measurements under various applied fields (see above). Interestingly, with increasing field, the lower *T* anomaly first shifts systematically towards higher temperature and it becomes very sharp signaling a possible first order type transition. At 1.55 T, both anomalies merge into one single anomaly which then starts to decrease in its absolute value and shifts towards lower temperature. Both elastic and inelastic neutron scattering experiments in high fields are essential to amass detailed information about the magnetic structure and magnetic excitations respectively. At 2 T, this feature takes a broadened hump like shape whose position moves towards higher temperature with increasing field strength. It is most likely caused by Zeeman effect. For high applied fields, as the temperature is lowered, *C/T* starts to deviate from -ln*T* dependence. Such deviation at very high fields was also seen in $Sr_3Ru_2O_7$ [38] and $CeAuSb_2$ [14]. Fig. 4(b) represents the entropy landscape showing *H -T* color map in the vicinity of the critical field region (1.4 T$\leq \mu_0 H \leq$ 1.9 T) with *S/T* as color scale (for raw entropy data at some specific fields see Fig. S5 in Supplementary Material [34]). The presented entropy landscape indicates that as the field is increased to the critical field region, *S/T* exhibits a peak centered in the critical field region. This observation strongly corroborates the existence of critical fluctuation suggesting the fact that $CePtIn_4$ is very close to the QCP under this conditions [39]. Similar feature in entropy landscape near the critical field region was also found in $Sr_3Ru_2O_7$ [11] due to the formation of nematic phase near the critical field. However, the later compound is a paramagnet under zero applied field.

The heat capacity data as presented in Fig. 4(a) spark the possibility of a tricritical point (TCP) at $T_N$ = 1.28 K and $\mu_0 H$ = 1.55 T which coincides with a magnetic triple point separating PM, AFM and intermediate metamagnetic (IMM)states. So, to unveil this apparent conjecture and to map the phase diagram around this TCP, we carried out magnetic field scans of heat capacity, *C(H)*, at different fixed temperatures. As can be seen from Fig. 4(c), C/T **(**H**)** displays quite astonishing behavior above and below this TCP (for *C(H)*at all temperature investigated see Fig. S6 in Supplementary Material [34]). At 1.3 K it shows one step like feature at around 1.5 T but, as the temperature is lowered, this step like feature becomes very sharp and then splits in two sharp transitions whose magnitudes gradually decreases with decreasing temperature. Appearance of two peaks in C/T **(**H**)** is consistent with the metamagnetic phase formation as probed by isothermal magnetization and MR measurements (see above). The second striking aspect of our *C(H)*data is the very narrow spread of applied fields at which the peaks are observed indicating the fact that within the accuracy of our measurements the AFM or MMT phase line becomes vertical in the H - T plane as $T_N \rightarrow 0$ leading to quantum critical end points (QCEP). The observation of anomalies in *C(H)*is quite rare in the existing literature and was seen in case of $Sr_3Ru_2O_7$ [11], $K_2PbCu(NO_2)_6$ [18] and $Yb_3Pt_4$ [35]

Summarizing the results obtained from above discussed experimental measurements, we constructed a magnetic phase diagram displayed as a color contour plot in Fig. 5 where $C(T)/T$ at different fixed fields (from Fig. 4a) is represented with the color scale. This phase diagram is quite unique as it manifests sequence of several magnetic phases of different nature. At low fields $T_N$ follows a power law (mean field like) form $T_N(H) = T_N(0)[1 - (H/H_c)^2]$ marked by the black dashed line in Fig. 5. This suggests that a conventional QCP would have been reached at $\mu_0 H_c \approx 2.4$ T. However, in reality, such a scenario is avoided and the observed second order phase line terminates at the TCP. Since the magnetic ordering is associated with a broken symmetry, the phase line must continue to $T_N = 0$ K as a line of first order transitions which will then terminate at QCEP. This is exactly what we observe in the present case. At TCP, the second order type phase line splits in to two almost vertical lines evincing first order like transitions which terminate at QCEPs. First order nature of the MMT was already confirmed by noticeable hysteresis in MR measurement (see above Fig 3d). Thus, it turns out that $CePtIn_4$ is very close to a field tuned QCP but in order to achieve a true QCP, we need to tune some other non-thermal parameter such as pressure or field angle. Another tempting aspect of the observed phase diagram is the possible relationship to the underlying electronic structure. Similar to the conclusions inferred for $CeRu_2Si_2$ [40], we anticipate that the magnetic field causes a continuous evolution of electronic structure where volume of one of the spin polarized Fermi surfaces is enhanced in increasing field while the other vanishes at a Lifshitz transition at metamagnetic field. Thus, Hall effect measurements across the MMT along with electronic band structure calculations are essential for $CePtIn_4$ to explore this exciting possibility.

## Conclusion

In conclusion, we investigated magnetic, electrical transport and thermodynamic properties of a novel antiferromagnetic compound $CePtIn_4$ to explore quantum criticality in this system. Magnetization measurements confirmed the occurrence of antiferromagnetic order and two subsequent metamagnetic transitions. Upon decreasing temperature differential susceptibility exhibits the tendency to power law divergence followed by saturation behavior, which clearly reflects the proximity of QCP. Metamagnetic transitions can also be easily discerned from the shape of magnetoresistance isotherms. Remarkably, MR below $T_N$ reveals a quite peculiar linear field dependence at high fields. All these observations, together with specific heat data collected in applied fields yielded a stunning phase diagram. It reveals that suppression of $T_N$ by applied field is terminated at TCP, which is also a triple point where antiferromagnetic, paramagnetic and intermediate metamagnetic states coexist. Beyond this point, two phase lines become first-order like and terminate at quantum critical end points. Thus, CePtIn4 turns out to be a unique case of strongly correlated electron system antiferromagnets with such perplexing phase diagram. Our description of $CePtIn_4$ indicating a substantial content of new mystifying physics in this compound opens new prospect for studies of quantum instabilities driven by magnetic field.

**Materials and Methods**

Single crystals of $CePtIn_4$ were grown using In flux method as outlined in our previous report [32]. We obtained plate like crystals grown along crystallographic *b*-axis. Even though, it crystallizes in the same crystal structure as reported by Bławat *et al.* [33], it has a larger unit cell volume as compared to the reported one. Surprisingly, these new single crystals show an enhancement in AFM ordering temperature with $T_N$ = 2.3 K. Magnetic measurements were performed in Quantum Design (QD) MPMS SQUID magnetometer expanded with iHe3 refrigerator. The electrical resistivity was measured over the temperature range 0.05 to 4 K and in applied magnetic fields up to 9 T using standard ac four-probe technique in QD PPMS (9T) dilution refrigerator assembly. Heat capacity measurements were performed in the range 0.1 to 4 K in fields up to 7 T using relaxation method in the very same dilution fridge system

**Acknowledgments**

The authors are indebted to Alicja Hackemer and Marek Daszkiewicz for assisting in synthesis of these crystals and the crystal structure characterization respectively. This work was supported by the National Science Centre (Poland) under research grant No. 2015/19/B/ST3/03158.

**Author contributions:** D.K. conceived the experiments and supervised the research. D.D., D.G. and PW performed the physical properties measurements. D.D. analyzed the experimental results. D.D. and D.K. wrote the manuscript with notable input from all the other authors.

**Competing interests:** The authors declare no competing interests.

**Figures**

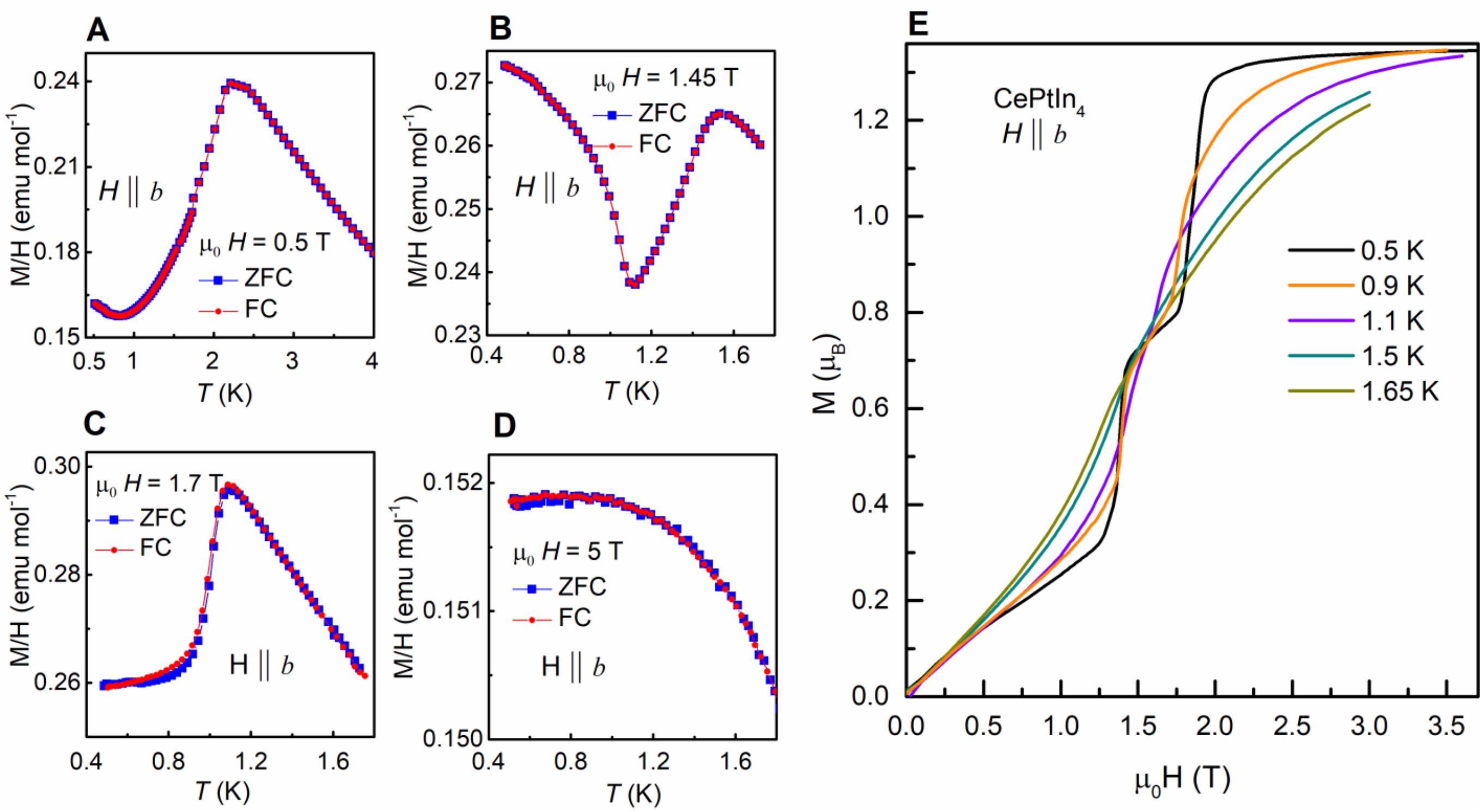

**Fig. 1. Bulk magnetic properties of $CePtIn_4$ single crystals.** (**A**)-(**D**) Temperature dependence of magnetization divided by the applied field, M=H, measured at different fields as cited in the figure. Applied fields are parallel to the crystallographic b-axis.(**E**) Field dependencies of magnetization at various fixed temperatures below $T_N$ as indicated.

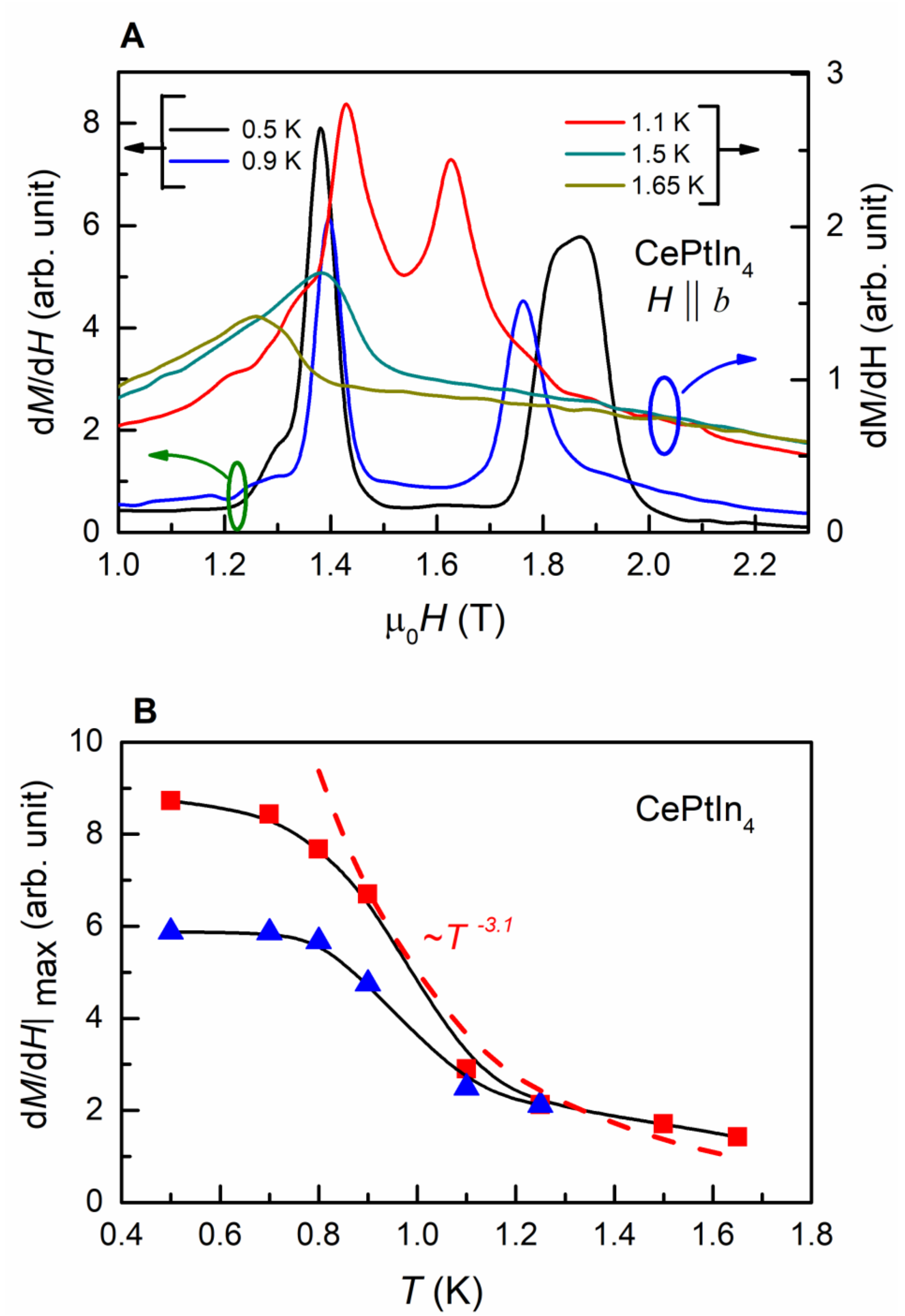

**Fig. 2. Field and temperature dependence of differential susceptibility and its maximum values for CePtIn4 single crystals .** (**A**) Field derivative of magnetization (dM/dH), also referred as differential susceptibility (see text), as a function of applied field. Sharp peaks in dM/dH vs H correspond to MMT in the system. (**B**) Temperature dependence of the maximum value of differential susceptibility, $dM/dH|_{max}$ derived from (**A**). Red square points correspond to the low field maxima whereas blue triangular points represent high field maxima. The black lines through these points are guide to the eyes. Red dashed line represents the fitting with power-law dependence as $\sim T^{-n}$.

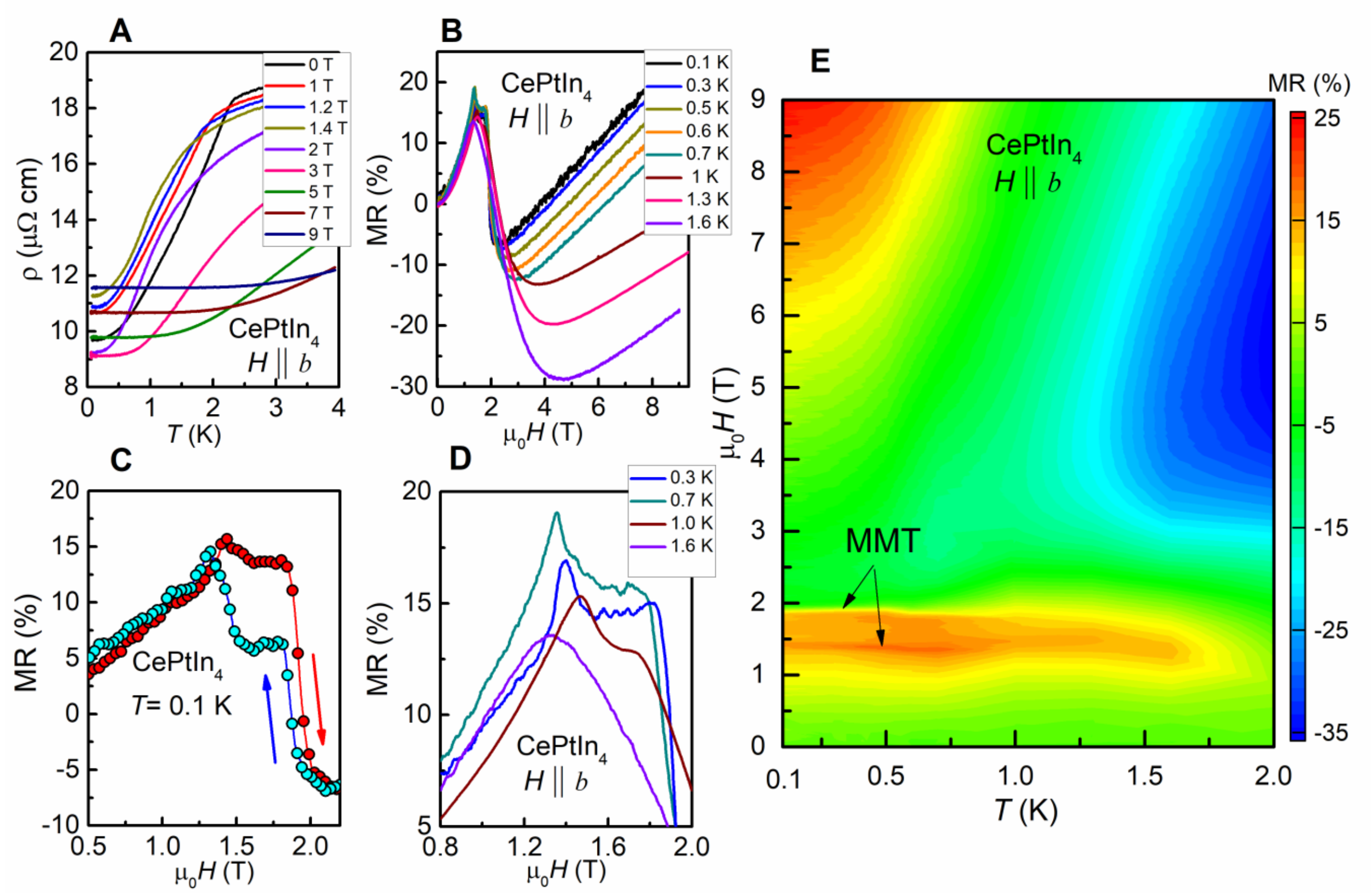

**Fig. 3. Magnetotransport properties of CePtIn₄ single crystals.** (**A**) Temperature dependence of electrical resistivity, ρ(T ), of single-crystalline CePtIn*4* measured with electric current flowing in ac-plane, in different fields applied along b-axis. (**B**) Magnetic field dependencies of the transverse MR of single-crystalline CePtIn*4* measured at several temperatures in the AFM state with electrical current flowing within ac-plane and magnetic field applied along the crystallographic b-axis. (**C**) magnetic field scan of MR at 0.1 K. A clear hysteresis between two MMTs, is indicative of first order nature of the transition. Vertical arrows indicate field up and down sweeps. (**D**) Magnified view of the MR data around MMT for few selective temperatures (see Fig 2 in Supplementary Information [36] for all temperatures). (**E**) Transverse MR of $CePtIn_4$ in AFM state depicted in a H - T color map plot with values of MR represented as the color scale. As highlighted by arrows, metamagnetic phase boundary is clearly visible in this plot.

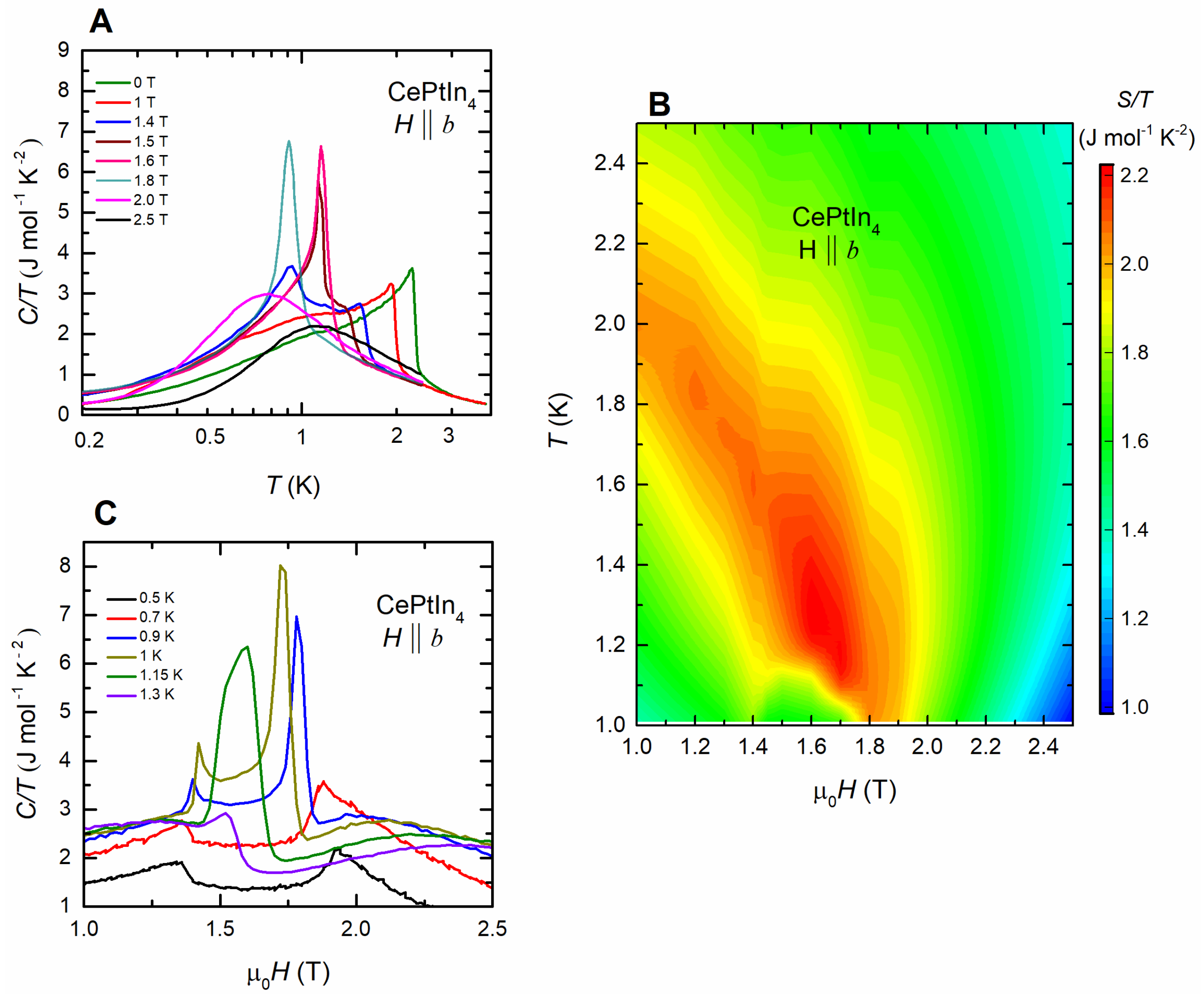

**Fig. 4. Heat capacity and entropy landscape of CePtIn4 single crystals.** (**A**) Temperature variation of the heat capacity over temperature ratio (*C/T*) measured for single-crystalline $CePtIn_4$ in various magnetic fields applied along the crystallographic *b*-axis. (**B**) Entropy landscape presented as *H* - *T* color map plot with entropy divided by temperature (*S/T*) represented by color scale (c) Field scan of heat capacity (presented as *C/T*) at different fixed temperatures.

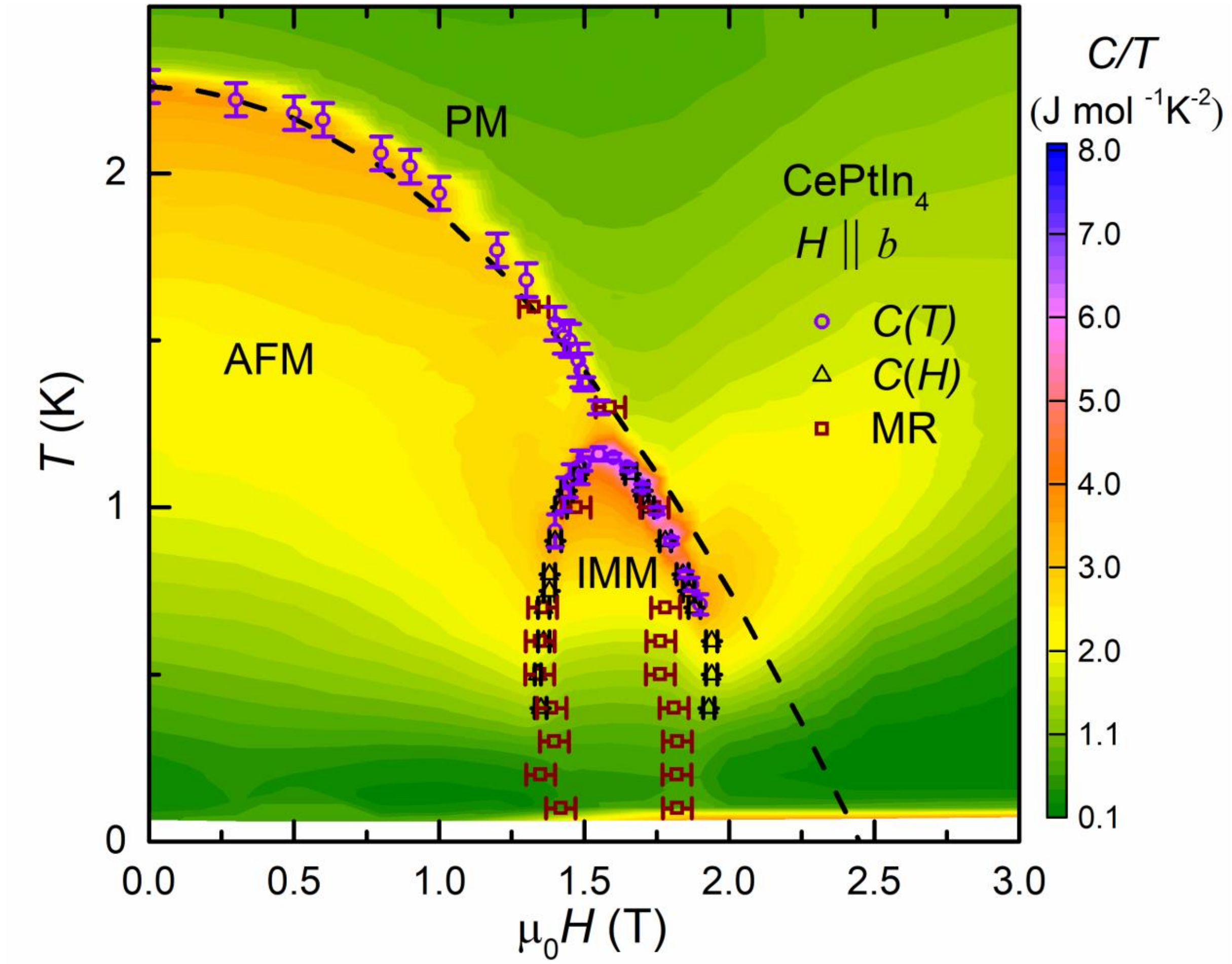

**Fig. 5. Magnetic phase diagram of CePtIn4 single crystals.** Magnetic phase diagram with applied field k*b*-axis presented as *H* - *T* color map plot with *C*=*T* (from Fig. 4a) represented as the color scale. Scattered points correspond to $T_N$ values obtained from different measurements as indicated in the figure. The dashed black line corresponds to mean field type dependence of $T_N$ (see text).

# Supplementary Information for

## Magnetic field driven quantum criticality in antiferromagnetic $CePtIn_4$

***Debarchan Das, Daniel Gnida, Piotr Wiśniewski, and Dariusz Kaczorowski***

***Prof. Dariusz Kaczorowski***
**Email: d.kaczorowski@intibs.pl**

In this supplementary information we present all experimental data collected on single crystalline $CePtIn_4$ sample.

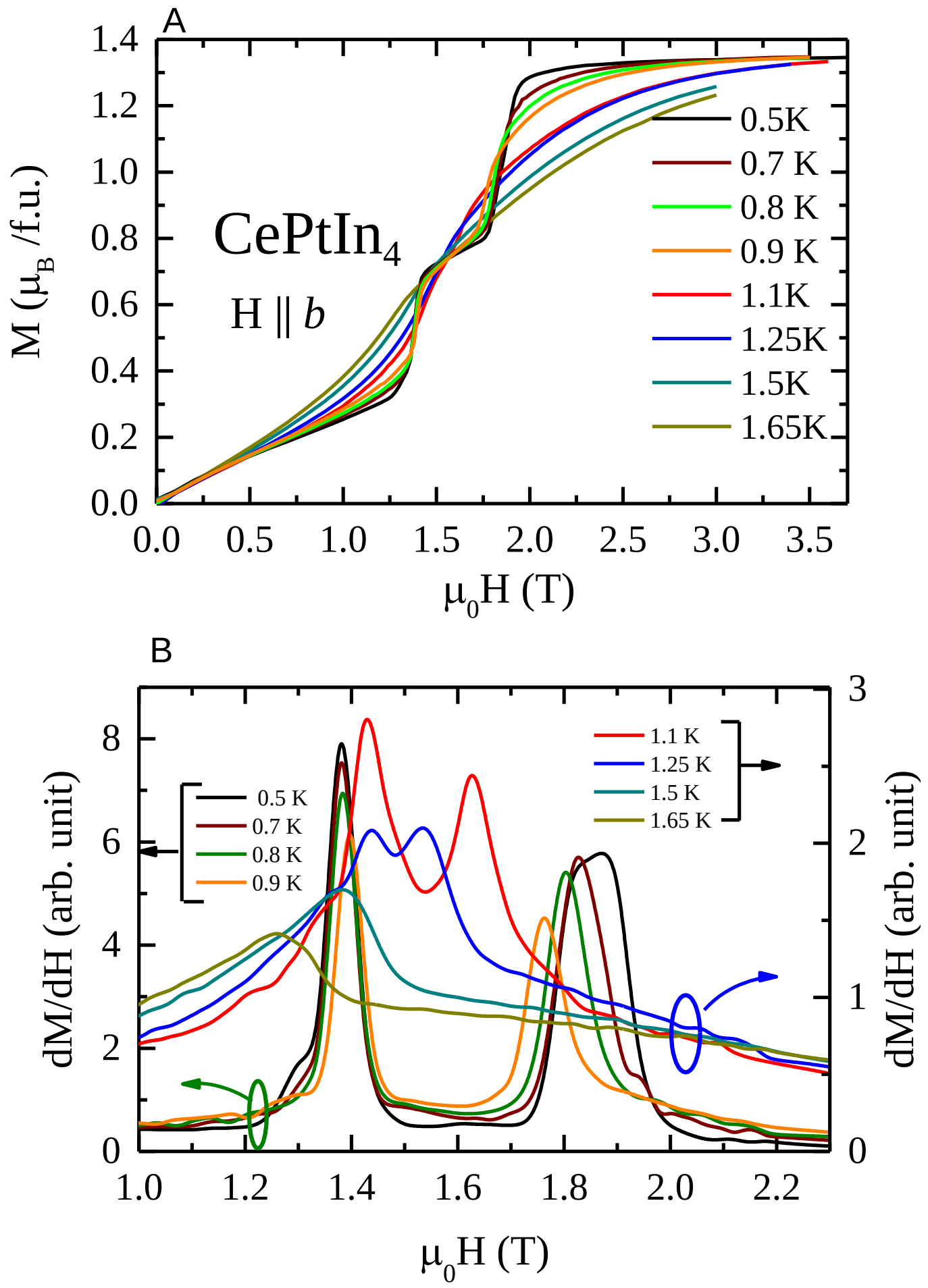

**Fig. S1.** (A)Field dependencies of magnetization at various fixed temperatures below $T_N$ as indicated. (B) Field derivative of magnetization(dM/dH)**,**also referred as differential susceptibility(see main text), as a function of applied field.

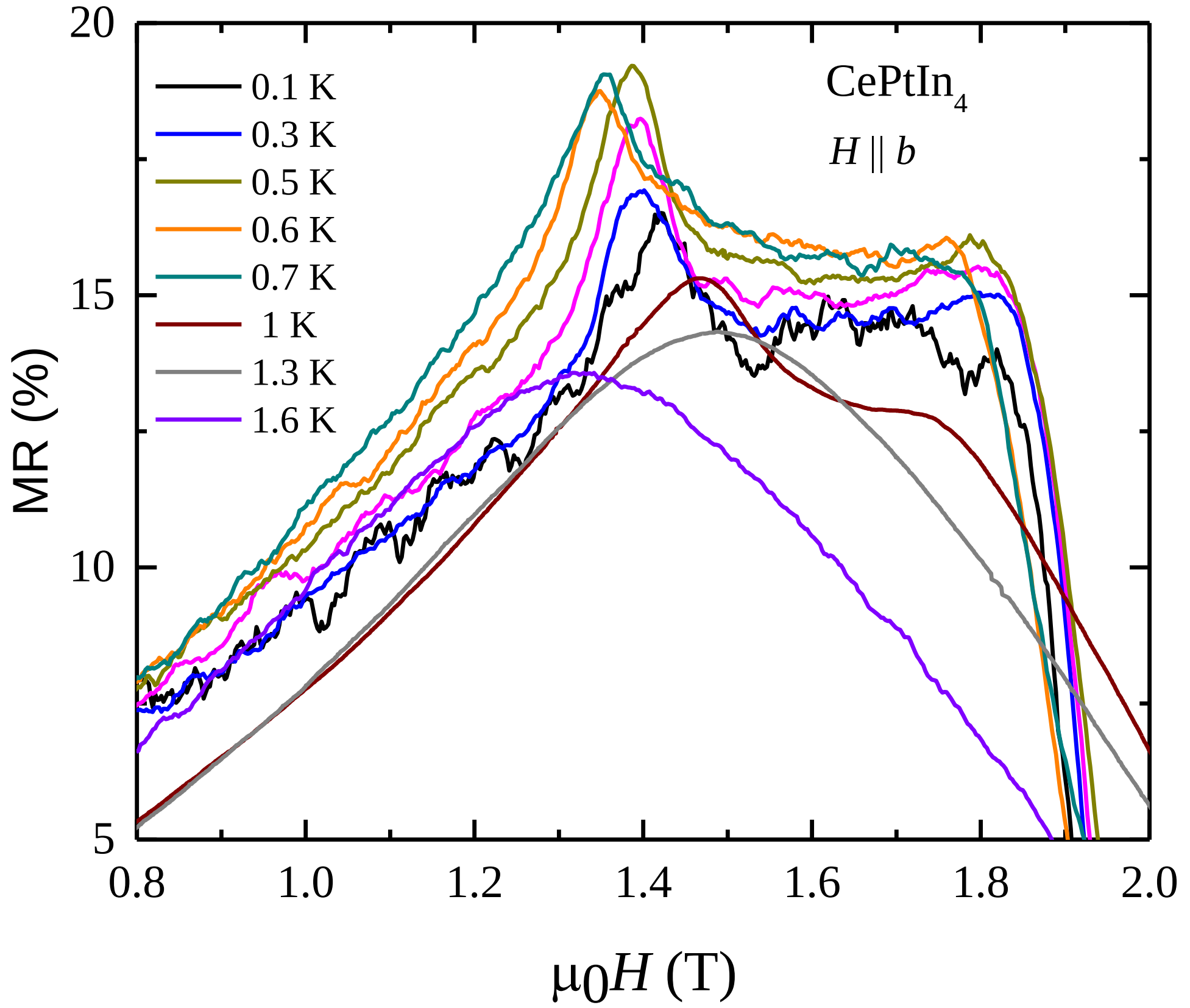

**Fig. S2.** Magnetic field dependencies of the transverse MR of single-crystalline CePtIn4 measured at several temperatures in the AFM state with electrical current flowing within **ac** plane and magnetic field applied along the crystallographic *b* axis.

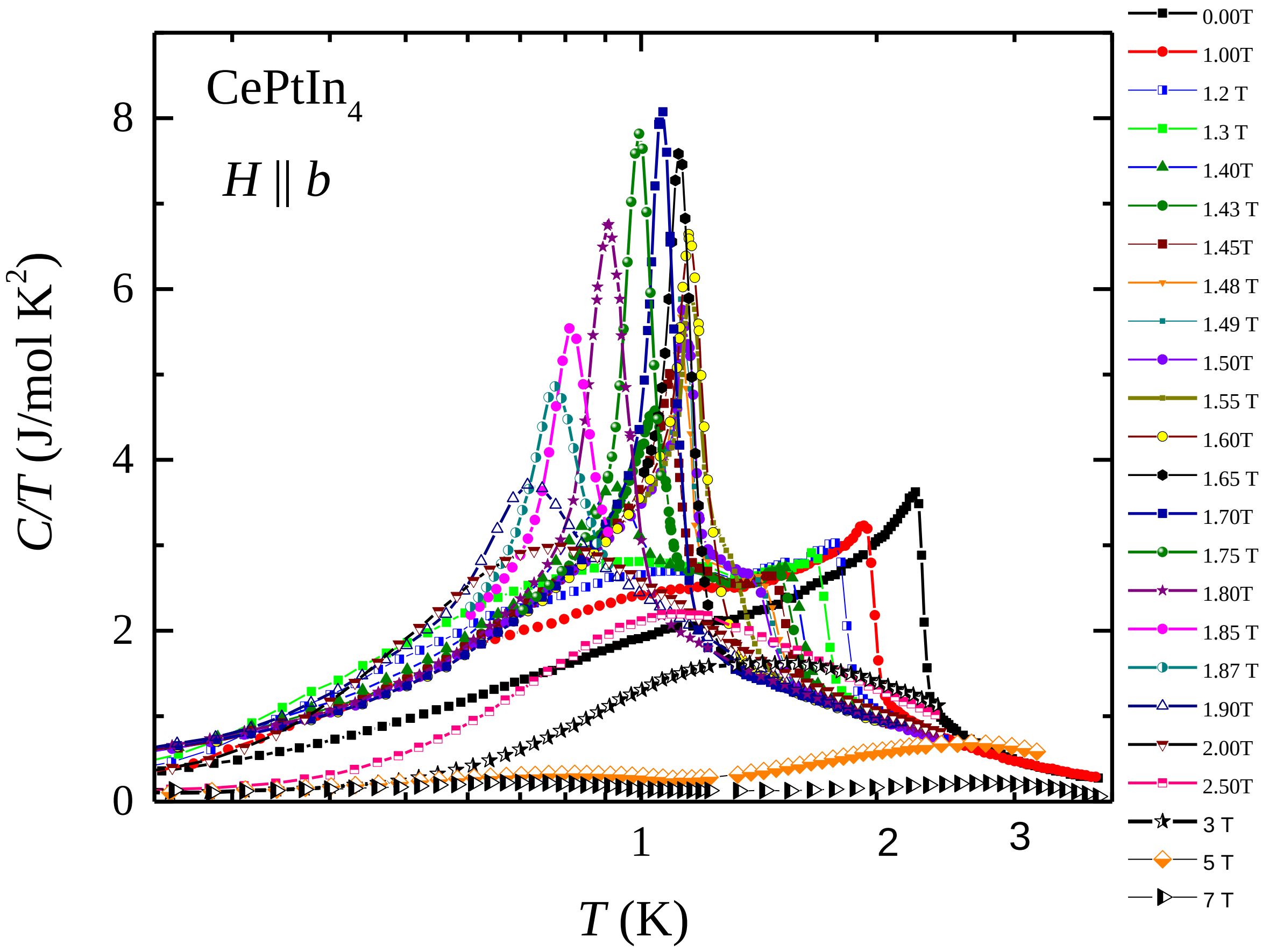

**Fig. S3.** Heat capacity over temperature ratio as a function of temperature at different applied fields up to 7T. These data were used to construct the phase diagram presented in Fig 5 in the main paper.

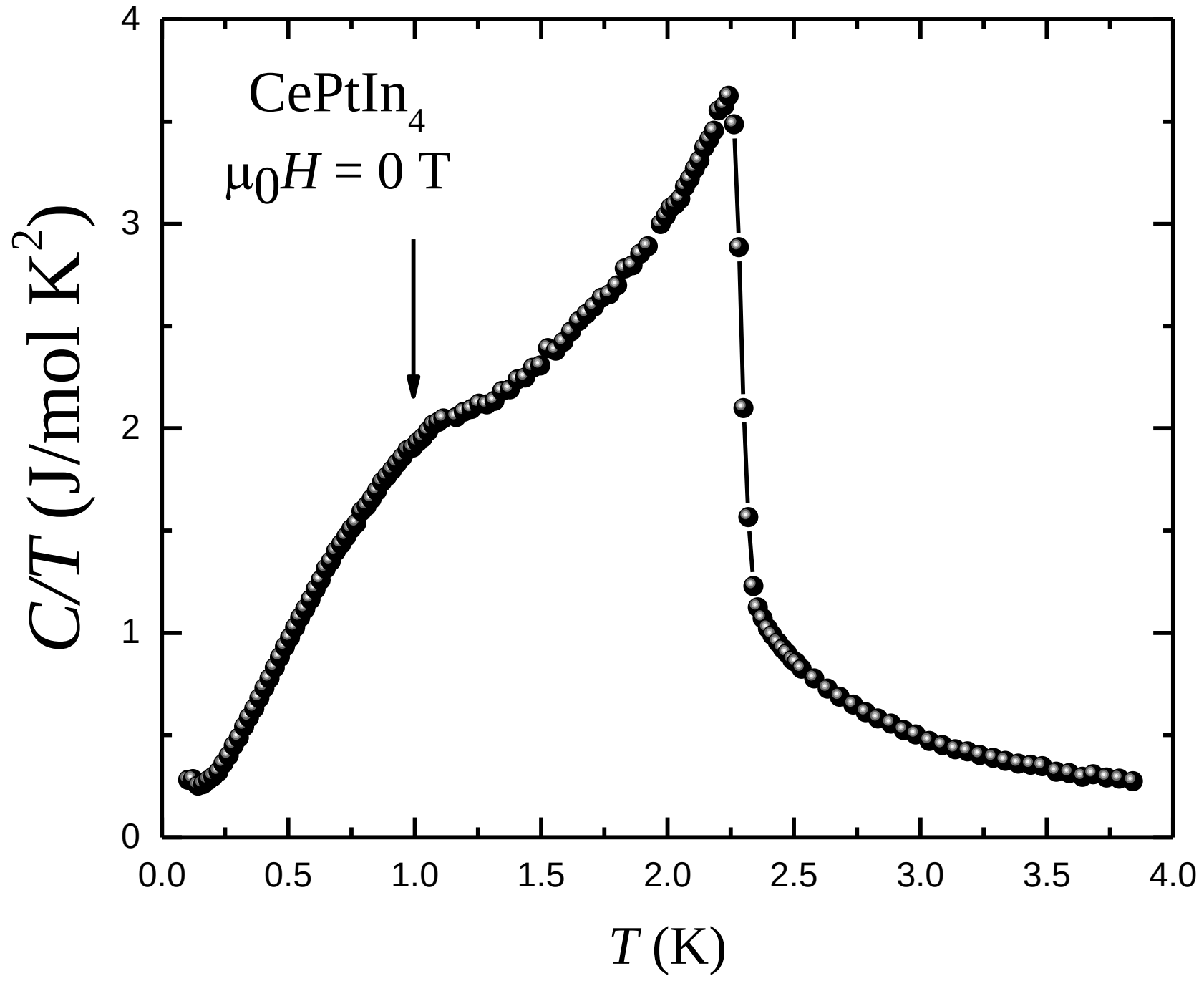

**Fig. S4.** Temperature dependence of heat capacity over temperature ratio highlighting sharp anomaly at $T_N$ = 2.3 K and another broad anomaly at ~1K (marked by vertical down arrow). The data above 0.4 K has also been presented in Ref [34] of the main paper [D. Das et al. Solid State Communications, **302**, 113717 (2019) ].

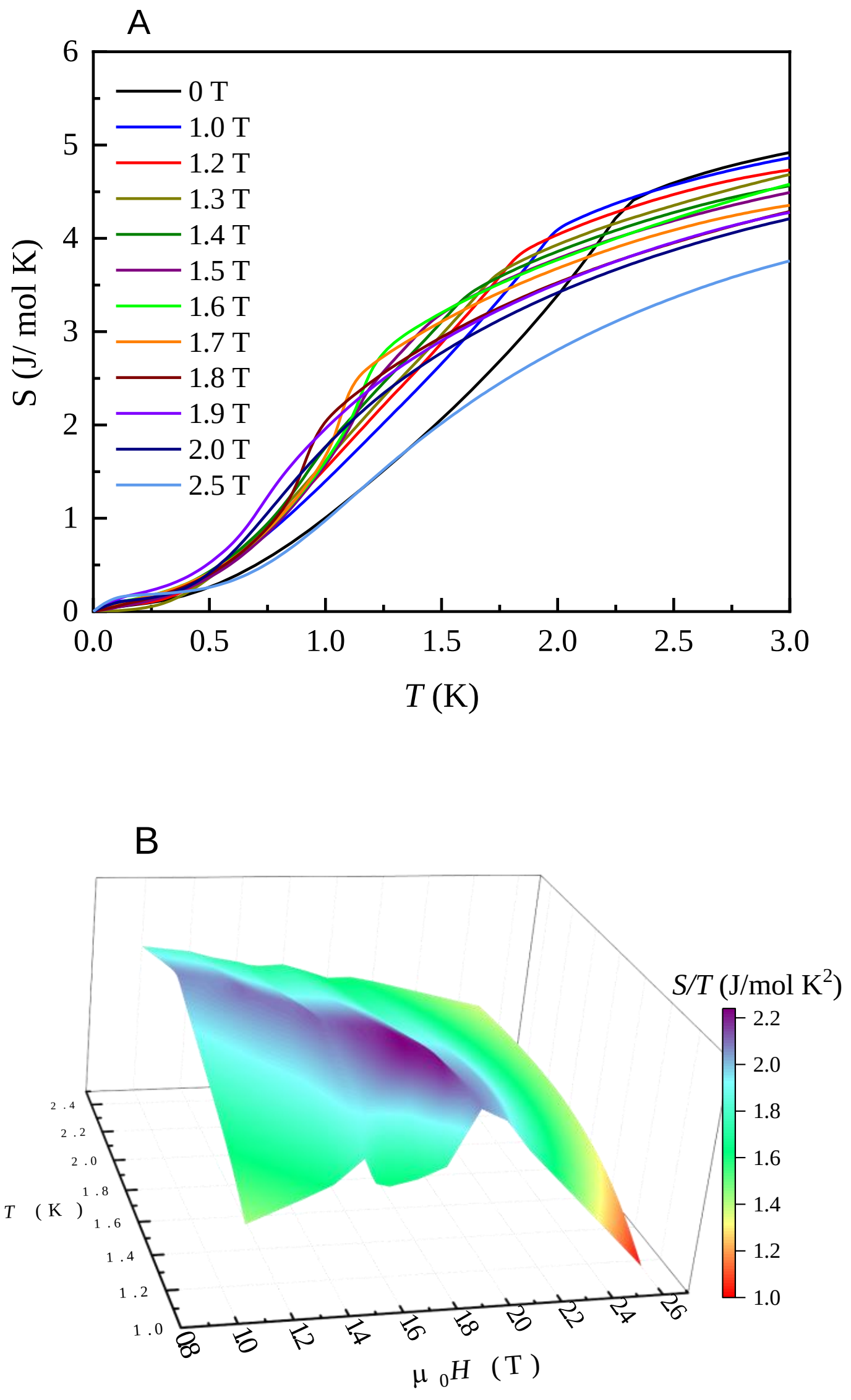

**Fig. S5. A.** Raw entropy data as a function of temperature for some selected applied fields. **B.** Entropy landscape presented as entropy divided by temperature (***S/T***) as a function of temperature and field. Around the TCP we observe a clear peak in *S/T*.

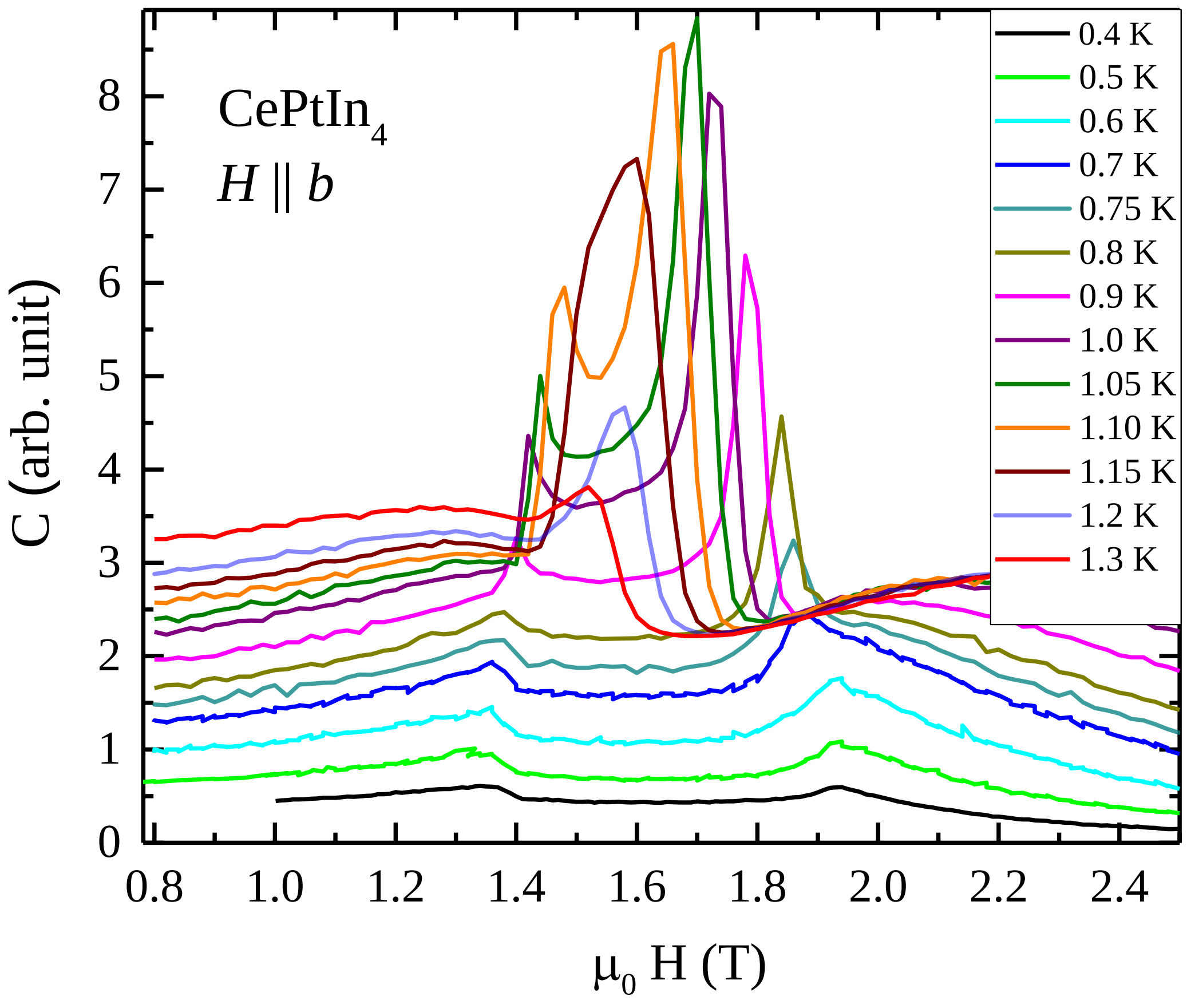

**Fig. S6**. All heat capacity field runs. These data were considered to plot the points mentioned as $C(\mu_0 H)$ in Fig 5 of the main paper.